\documentstyle[pre,aps,preprint,epsfig,floats]{revtex}
\tightenlines
\begin{document}
\draft

\title{Current Reversals in Chaotic Ratchets}
\author{Jos\'e L. Mateos}
\address{Instituto de F\'{\i}sica, 
Universidad Nacional Aut\'onoma de M\'exico,\\
Apartado Postal 20-364, M\'exico D.F. 01000, M\'exico}

\maketitle
\begin{abstract}
The problem of the classical deterministic dynamics of a 
particle in a periodic asymmetric potential of the ratchet type is
addressed. When the inertial term is taken into account, the
dynamics can be chaotic and modify the transport properties. 
By a comparison between the bifurcation diagram and the current, 
we identify the origin of the current reversal as a bifurcation 
from a chaotic to a periodic regime. Close to this bifurcation, 
we observed trajectories revealing intermittent chaos and anomalous 
deterministic diffusion. We extend our previous analysis of this 
problem to include multiple current reversal and the orbits in 
phase space.  
\end{abstract}

\vspace{0.8cm}

\section{Introduction}
There is an increasing interest in recent years in the
study of the transport properties of nonlinear systems that can 
extract usable work from unbiased nonequilibrium fluctuations. 
These, so called ratchet systems, can be modeled, for instance,
by considering a Brownian particle in a periodic asymmetric 
potential and acted upon by an external time-dependent force 
of zero average. For recent reviews see \cite{han,ast,jul,bie}. 
This recent burst of work is motivated in part by the challenge
to explain the unidirectional transport of molecular motors 
in the biological realm \cite{bie}. Another source of 
motivation arises from the potential for new methods of 
separation of particles, polyelectrolytes and macromolecules 
\cite{rou,bie2,der,van,san,bie3}, and more recently in the 
recognition of the ``ratchet effect'' in the quantum domain 
\cite{rei,lin,lin2,lin3,zap,zap2}. 

In order to understand the generation of unidirectional motion
from nonequilibrium fluctuations, several models have been used. 
In Ref. \cite{han}, there is a classification of different 
types of ratchet systems; among them we can mention the 
``Rocking Ratchets'', in which the particles move in an asymmetric 
periodic potential subject to spatially uniform, time-periodic 
deterministic forces of zero average. Most of the models, 
so far, deal with the overdamped case in which the inertial 
term due to the finite mass of the particle is neglected. 
However, more recently, this oversimplification was 
overcome by treating properly the effect of finite mass
\cite{jun,mat,por,por2,fla,bar}. 

In particular, in two recent papers \cite{jun,mat}, the authors 
studied the effect of finite inertia in a deterministically 
rocked, periodic ratchet potential.
They consider the deterministic case in which noise is 
absent. The inertial term allows the possibility of 
having both regular and chaotic dynamics, and this 
deterministically induced chaos can mimic the role of 
noise. They showed that the system can exhibit a current 
flow in either direction, presenting multiple current 
reversals as the amplitude of the external force is varied. 

In Ref. \cite{mat}, the role of the chaotic dynamics in the 
current was analyzed in detail, establishing for the first
time a close connection between the current and the
bifurcation diagram when a control parameter of the 
model is varied. In this paper we elaborate on this 
idea by studying the multiple current reversals and
the orbits in phase space. 

The outline of the paper is as follows: in the next section 
we introduce the equations of motion that define the
model, and in the next section we present the numerical
results. We end with some concluding remarks in the 
last section.

\section{The ratchet potential model}

Let us consider the one-dimensional problem of a particle 
driven by a periodic time-dependent external force, under 
the influence of an asymmetric periodic potential of the 
ratchet type. The time average 
of the external force is zero. Here, we do not take into account 
any kind of noise, and thus the dynamics is deterministic. 
The equation of motion is given by

\begin{equation}
m\ddot x + \gamma \dot x + {\frac{dV(x)}{dx}} = F_0 \cos(\omega_D t),
\end{equation}

\noindent where $m$ is the mass of the particle, 
$\gamma$ is the friction coefficient,
$V(x)$ is the external asymmetric periodic potential, $F_0$ is the amplitude
of the external force and $\omega_D$ is the frequency of the external
driving force. The ratchet potential is given by 

\begin{equation}
V(x) = V_1 - V_0 \sin {\frac{2\pi (x-x_0)}{L}} - {\frac{V_0}{4}}
\sin {\frac{4\pi (x-x_0)}{L}},
\end{equation}

\noindent where $L$ is the periodicity of the potential, 
$V_0 $ is the amplitude, and $V_1$ is an arbitrary constant. The potential
is shifted by an amount $x_0$ in order that the minimum of the potential
is located at the origin.

Let us define the following dimensionless units: ${x^{\prime }=x/L}$,
$x_{0}^{\prime }=x_{0}/L$, $t^{\prime }=\omega _{0}t$,
$w=\omega _{D}/\omega_{0}$, $b=\gamma /m\omega _{0}$ and
$a=F_{0}/mL\omega _{0}^{2}$. 
Here, the frequency $\omega _{0}$ is given by $\omega _{0}^{2}=4\pi
^{2}V_{0}\delta /mL^{2}$ and $\delta $ is defined by $\delta =\sin (2\pi
|x_{0}^{\prime }|)+\sin (4\pi |x_{0}^{\prime }|)$. 

The frequency $\omega _{0}$ is the frequency of the linearized motion 
around the minima of the potential, thus we are scaling the time with 
the natural period of motion 
$\tau _{0}=2\pi /\omega _{0}$. The dimensionless equation
of motion, after renaming the variables again without the primes, becomes 

\begin{equation}
\ddot{x}+b\dot{x}+{\frac{dV(x)}{dx}}=a\cos (wt),   
\end{equation}

\noindent where the dimensionless potential can be written as

\begin{equation}
V(x) = C - {\frac{1}{4\pi^{2}\delta}} \left [\sin 2\pi (x-x_{0}) +
{\frac{1}{4}} \sin 4\pi (x-x_{0}) \right ]
\end{equation}

\noindent and is depicted in Fig. 1. The constant $C$ is
such that $V(0)=0$, and is given by
$C = -(\sin 2\pi x_{0} + 0.25 \sin 4\pi x_{0})/4\pi^{2}\delta$.
In this case, $x_{0}\simeq -0.19$, $\delta \simeq 1.6$ and 
$C\simeq 0.0173$.

In the equation of motion Eq. (3) there are three
dimensionless parameters: $a$, $b$ and $w$, defined 
above in terms of
physical quantities. We will vary the parameters in order to 
understand the role of each in the dynamics. The parameter
$a=F_{0}/mL\omega _{0}^{2}$ is the ratio of the amplitude of the
external force and the force due to the potential $V(x)$. This 
can be seen more clearly using the expression for $\omega _{0}^{2}$
in terms of the parameters of the potential. In this case, the
ratio becomes $a={\frac{1}{4\pi^{2}\delta}} F_{0}/(V_{0}/L)$, that is, 
except for a constant factor, $a$ is the ratio of $F_{0}$ and the 
force $V_{0}/L$, where $V_{0}$ is the amplitude and $L$ the 
periodicity of the potential (see Eq. (2)).

The parameter $b$ is simply the dimensionless friction coefficient, 
and $w$ is the ratio of the driving frequency of the external force 
and $\omega_{0}$. We will discuss in more detail these 
parameters in the next section.

The extended phase space in which the dynamics is taking place is
three-dimensional, since we are dealing with an inhomogeneous
differential equation with an explicit time dependence. This equation
can be written as a three-dimensional dynamical system, that we solve 
numerically, using the fourth-order Runge-Kutta algorithm.
The equation of motion Eq. (3) is nonlinear and thus allows the
possibility of periodic and chaotic orbits. If the inertial term 
associated with the second derivative $\ddot{x}$ were absent, 
then the dynamical system could not be chaotic.

The main motivation behind this work is to study in detail the origin of
the current reversal in a chaotically deterministic rocked ratchet as 
found in \cite{mat}. In order to do so, we have to study first the 
current $J$ itself, that we define as the time average of the 
average velocity over an ensemble of initial conditions. Therefore,
the current involves two different averages: the first average is
over $M$ initial conditions, that we take equally distributed in
space, centered around the origin and with an initial velocity
equal to zero. For a fixed time, say $t_j$, we obtain an average
velocity, that we denoted as $v_j$, and is given by 

\begin{equation}
v_j = {\frac{1}{M}}\sum\limits_{i=1}^M {\dot{x_i}}(t_j).
\end{equation}

The second average is a time average; since we take a discrete time
for the numerical solution of the equation of motion, we have a
discrete finite set of $N$ different times $t_j$; 
then the current can be defined as 

\begin{equation}
J={\frac{1}{N}}\sum\limits_{j=1}^N {v_j}.
\end{equation}

This quantity is a single number for a fixed set of parameters 
${a,b,w}$.

Besides the orbits in the extended phase space, 
we can obtain the Poincar\'e section,
using as a stroboscopic time the period of oscillation of the
external force. With the aid of Poincar\'e sections we can
distinguish between periodic and chaotic orbits, and we can obtain
a bifurcation diagram as a function of the parameter $a$. As was
shown in \cite{mat}, there is a connection between the bifurcation 
diagram and the current. 

\section{Numerical results}

Using the definition of the current $J$ given in the previous
section, we calculate $J$ fixing the parameters $b=0.1$ 
and $w=0.67$ and varying the parameter $a$. The current
shows, as stressed before \cite{jun,mat}, multiple
current reversals and a complex variation with $a$, as shown 
in Fig. 2b. We can observe strong fluctuations as well as 
portions where the current is approximately 
constant. The challenge here is to explain this high 
variability in the current with the aid of what we know from
the nonlinear chaotic dynamics of the system.

Associated with this current, there is a correspondent 
bifurcation diagram as a function of $a$, as depicted in
Fig. 2a. The complexity in this diagram is a consequence
of the richness in the dynamics of the particle in the 
non-linear ratchet potential. We notice that the bifurcation
diagram  for the ratchet is qualitatively similar to the 
bifurcation diagram of a harmonically forced pendulum with
friction \cite{bak}. We can imagine the problem of the pendulum
as a particle in a {\it symmetric} periodic potential that varies
in time. In this sense, our ratchet problem is an asymmetric 
generalization of the pendulum where a spatial symmetry 
breaking occurs. There is a recent work \cite{fla} that studied
the pendulum and the ratchet in the context of symmetry
breaking. 

In order to understand the first part of the current, let us analyze
the case of small values of $a$, where we chose $b=0.1$ and 
$w=0.67$. In Fig. 3a we show the current as a function of $a$,
and in Fig. 3b we depict the bifurcation diagram in the same 
range of $a$. Let us imagine that an ensemble
of particles are initially located at the minimum of the ratchet 
potential around the origin, and that all these particles have an 
initial velocity equal to zero. For $a=0$, we have no external force
and thus, all these particles remain in the minimum around the
origin and therefore the current is zero. For very small values of
$a$, we still have a zero current, since the particles have friction
and tend to oscillate in this minimum. However, there is a critical
value of $a$ for which the particles start to overcome the potential
barriers around the minimum and transport along the ratchet
potential in a periodic or chaotic way. This critical value can be 
calculated as follows: remember that $a$ is a dimensionless 
quantity defined as $a={\frac{1}{4\pi^{2}\delta}} F_{0}/(V_{0}/L)$.
Here, $F_{0}$ is the amplitude of the external force and 
$V_{0}/L$ is the order of magnitude of the force exerted by the 
potential. Thus, we expect the current $J$ be different from zero 
when $F_{0}$ is on the order of $V_{0}/L$, that is, 
$F_{0} \sim V_{0}/L$. In this case, the critical value of $a$ is
$a_c\sim 1/4\pi^{2}\delta$, since $F_{0}/(V_{0}/L) \sim 1$. 
Using the value of $\delta \simeq 1.6$ we obtain $a_c \sim 0.1$,
which is on the order of magnitude of the values that we obtain 
numerically. Above this value, the current starts to grow since more 
and more particles contribute to the current.

At the beginning, the current is dominated by transport due to 
periodic orbits, but for larger values of $a$, some of the orbits in 
the ensemble become chaotic and the transport is not as efficient as
before, resulting in a current that starts to oscillate erratically. In fact,
in this region, there exist the possibility of coexistence of multiple 
attractors in the phase space. For example, in Fig. 3a, we have two
coexistent attractors: a periodic and a chaotic one, around 
$a=0.067$. In this case, depending on the initial conditions, some 
orbits in the ensemble can end up in the periodic attractor, and the
rest in the chaotic attractor.

For values of $a$ even larger, all the orbits enter a chaotic region 
through a period-doubling bifurcation, and the current starts to 
decrease inside this chaotic band. Finally, exactly at the bifurcation
point where a periodic window opens, the current drops to zero and
becomes negative in a very abrupt way \cite{mat}.

Let us focus first on the range of the control parameter where the first
current reversal takes place. This occurs around $a\simeq 0.08$ as
shown in Fig. 3. We can observe a period-doubling route to chaos
and after a chaotic region, there is a saddle-node bifurcation taking
place at the critical value $a_c\simeq 0.08092844$. It is precisely at 
this bifurcation point that the current reversal occurs. 
After this bifurcation, a periodic window emerges, with an 
orbit of period four. In Figs. 3a,b we are analyzing only a short 
range of values of $a$, where the first current reversal takes place.
If we vary $a$ further, we can obtain multiple current 
reversals, as shown in Fig. 2b.

In order to understand in more detail the nature of the current reversal, 
let us look at the orbits just before and after the transition. The 
reversal occurs at the critical value $a_c\simeq  0.08092844$. 
If $a$ is below this critical value $a_c$, say $a = 0.07$, 
then the orbit is periodic, with 
period two. For this case we depict, in Fig. 4a, the
position of the particle as a function of time. We notice a period-two 
orbit, as can be distinguish in the bifurcation diagram for $a = 0.07$. 
This orbit transport particles to the positive direction and the 
corresponding velocity is a periodic function of time of period two,
as shown in Fig. 4b. The phase space for this orbit is depicted in 
Fig. 4c. We notice that the particle oscillates for a while around the
minima of the ratchet potential, before moving to the next one. The
spatial asymmetry of the potential is apparent in this orbit in 
phase space. 

In Fig. 5a we show again the position as a function of time for 
$a = 0.081$, which is just above the critical value $a_c$. In this case, 
we observe a period-four orbit, that corresponds to the periodic
window in the bifurcation diagram in Fig. 3a. This orbit is
such that the particle is ``climbing'' in the negative direction, 
that is, in the direction in which the slope of the potential is higher.
We notice that there is a qualitative difference
between the periodic orbit that transport particles to the positive 
direction and the periodic orbit that transport particles to the 
negative direction: in the latter case, the particle requires twice
the time than in the former case, to advances one well in the 
ratchet potential. A closer look at the trajectory in Fig. 5a reveals 
the ``trick'' that the particle uses to navigate in the negative 
direction: in order to advance one step to the left, it moves first 
one step to the right and then two steps to the left. The net result is 
a negative current.

The period-four orbit is apparent in Fig. 5b, where we show the 
velocity as a function of time. In Fig. 5c we depict the corresponding
phase space for this case. The transporting orbit is more elaborate
because it involves motion to the positive and negative directions, 
as well as oscillations around the minima.
 
In Fig. 6a, we show a typical trajectory for $a$ just below $a_c$. 
The trajectory is chaotic and the corresponding chaotic attractor is 
depicted in Fig. 7. In this case, the particle starts at the origin with 
no velocity; it jumps from one well in the ratchet potential to another 
well to the right or to the left in a chaotic way. The particle gets 
trapped oscillating for a while in a minimum (sticking mode), as 
is indicated by the integer values of $x$ in the ordinate, and 
suddenly starts a running mode with average constant velocity in 
the negative direction. In terms of the velocity, these running 
modes, as the one depicted in Fig. 5a, correspond  to periodic 
motion. This can be seen more clearly in Fig. 6b, where we plot the
velocity as a function of time in the same range of values as the
orbit in Fig. 6a. In Fig. 6c we show the corresponding phase space. 

The phenomenology can be described as follows. For values of 
$a$ above $a_c$, as in Fig. 5a, the attractor is a periodic orbit. 
For $a$ slightly less than $a_c$ there are long stretches of time 
(running or laminar modes) during which the orbit appears to be 
periodic and closely resembles the orbit 
for $a>a_c$, but this regular (approximately periodic) behavior is 
intermittently interrupted by finite duration ``bursts'' in which the
orbit behaves in a chaotic manner. The net result in the velocity 
is a set of periodic stretches of time interrupted by burst of 
chaotic motion, signaling precisely the phenomenon of intermittency 
\cite{ott}. As $a$ approach $a_c$ from below, the duration of the
running modes in the negative direction increases, until the duration
diverges at $a=a_c$, where the trajectory becomes truly periodic.
 
To complete this picture, in Fig. 7, we show two attractors: 
(1) the chaotic attractor for $a=0.08092$, just below $a_c$, 
corresponding to the trajectory in Fig. 6a, and (2) the 
period-4 attractor for $a=0.08093$, corresponding to the 
trajectory in Fig. 5a. This periodic attractor consist of four 
points in phase space, which are located at the center of
the open circles. We obtain these attractors confining the 
dynamics in $x$ between $-0.5$ and $0.5$, that is, we used
the periodicity of the potential $V(x+1) = V(x)$, to map the
points in the $x$ axis modulo 1. Thus, even though the
trajectory transport particles to infinity, when we confine the
dynamics, the chaotic structure of the attractor is apparent.
As $a$ approaches $a_c$ from below, the dynamics in 
the attractor becomes intermittent, spending most of the
time in the vicinity of the period-four attractor, and 
suddenly ``jumping'' in a chaotic way for some time, and then 
returning close to the period-four attractor again, and so on. 
In terms of the velocity, the result is an intermittent time 
series as the one depicted in Fig. 6b.

In order to characterize the deterministic diffusion in this regime, 
we calculate the mean square displacement 
$\langle (x - \langle x \rangle )^2 \rangle$ as a function of time. 
We obtain numerically that 
$\langle (x - \langle x \rangle )^2 \rangle \sim t^{\alpha}$, where the
exponent $\alpha \simeq 3/2$. This is a signature of anomalous
deterministic diffusion, in which 
$\langle (x - \langle x \rangle )^2 \rangle$ grows faster
than linear, that is, $\alpha > 1$ (superdiffusion). Normal 
deterministic diffusion corresponds to $\alpha = 1$.
In contrast, the trajectories in Figs. 4a and 5a transport 
particles in a ballistic way, with $\alpha = 2$.
The relationship between anomalous deterministic diffusion and
intermittent chaos has been explored recently, together with 
the connection with L\'evy flights \cite{shl}. The character
of the trajectories, as the one in Fig. 6a, remains to 
be analyzed more carefully in order to determine if they 
correspond to L\'evy flights.
 
\section{Concluding remarks}

In summary, we have studied the chaotic dynamics of 
a particle in a ratchet potential under the influence of an
external periodic force. We establish a connection 
between the bifurcation diagram and the current and
identify the mechanism by which the current 
reversal in deterministic ratchets arises: it corresponds to 
a bifurcation from a chaotic to a periodic regime. Near this 
bifurcation, the chaotic trajectories exhibit intermittent 
chaos and the transport arises through deterministic 
anomalous diffusion with an exponent greater than one.
The richness and the complexity of the bifurcation 
diagram and the asociated current, urge us to study
their connection in more detail in the near future.

\vfill\eject

\begin{figure}[htb]
\centerline{\epsfig{file=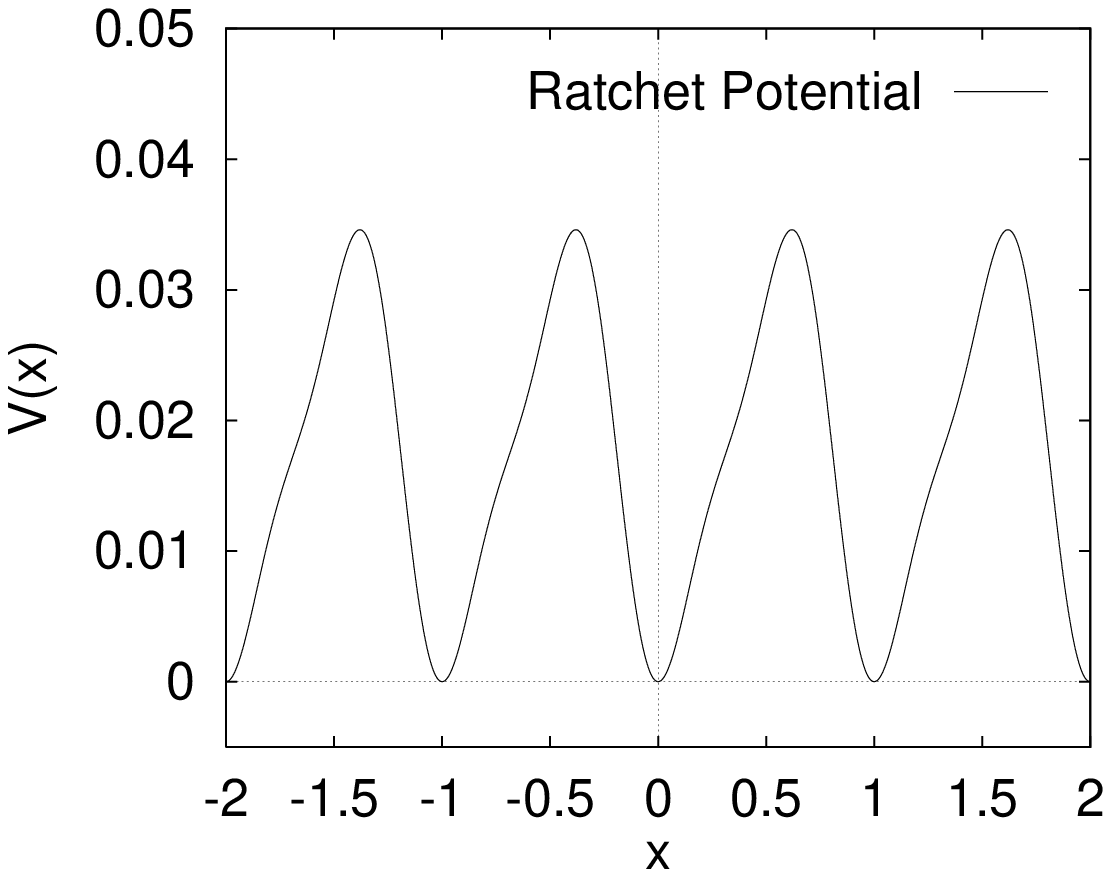,width=7.0cm}}
\caption{The dimensionless ratchet periodic potential $V(x)$.}
\label{fig1}
\end{figure}

\vfill\eject

\begin{figure}[htb]
\centerline{\epsfig{file=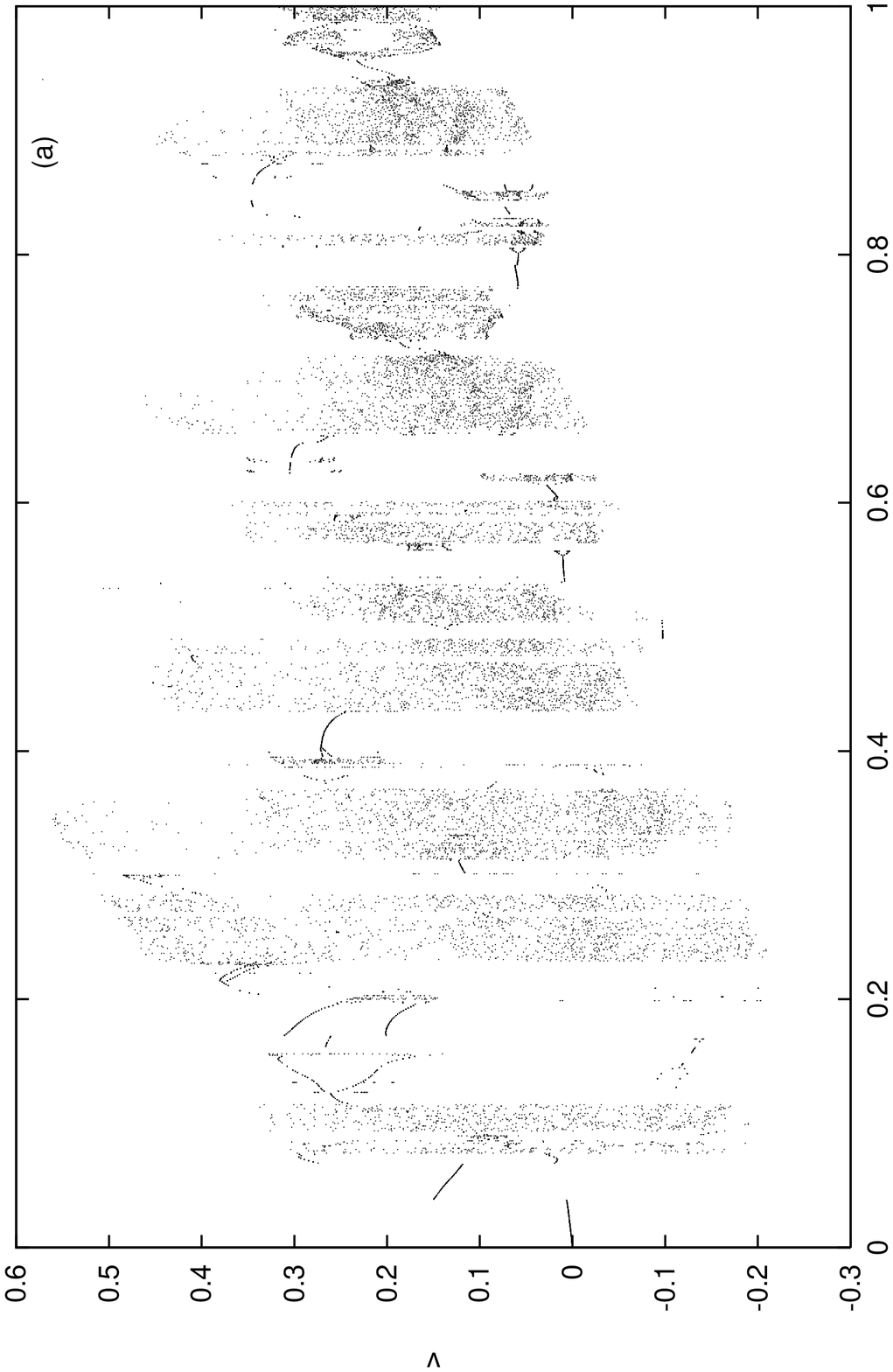,width=8.6cm}}
\label{f2a}
\end{figure}

\begin{figure}[htb]
\centerline{\epsfig{file=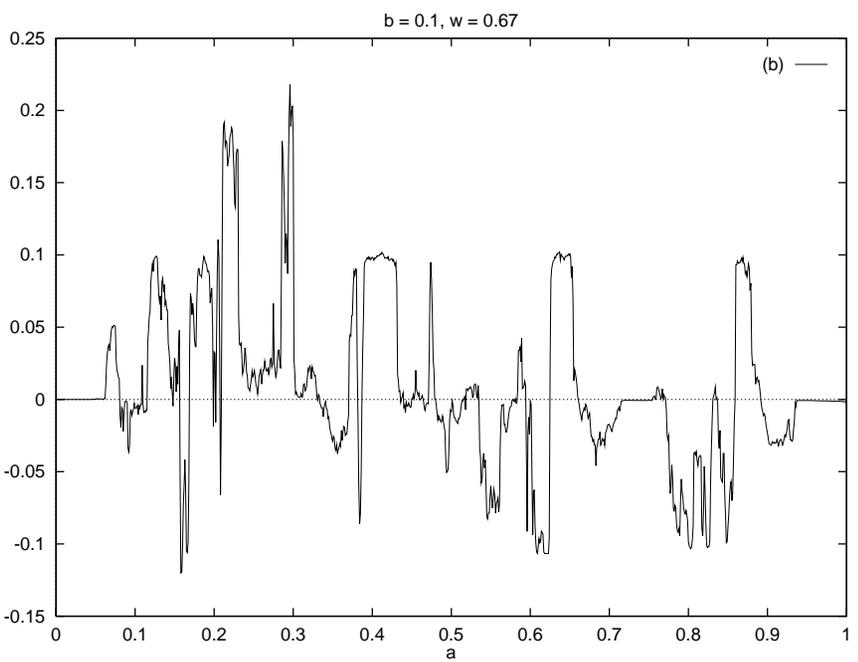,width=8.6cm}}
\caption{For $b=0.1$ and $w=0.67$ we show: (a) The bifurcation
diagram as a function of $a$, and (b) The current $J$ as a
function of $a$. We can see multiple current reversals.}
\label{fig2b}
\end{figure}

\vfill\eject

\begin{figure}[htb]
\centerline{\epsfig{file=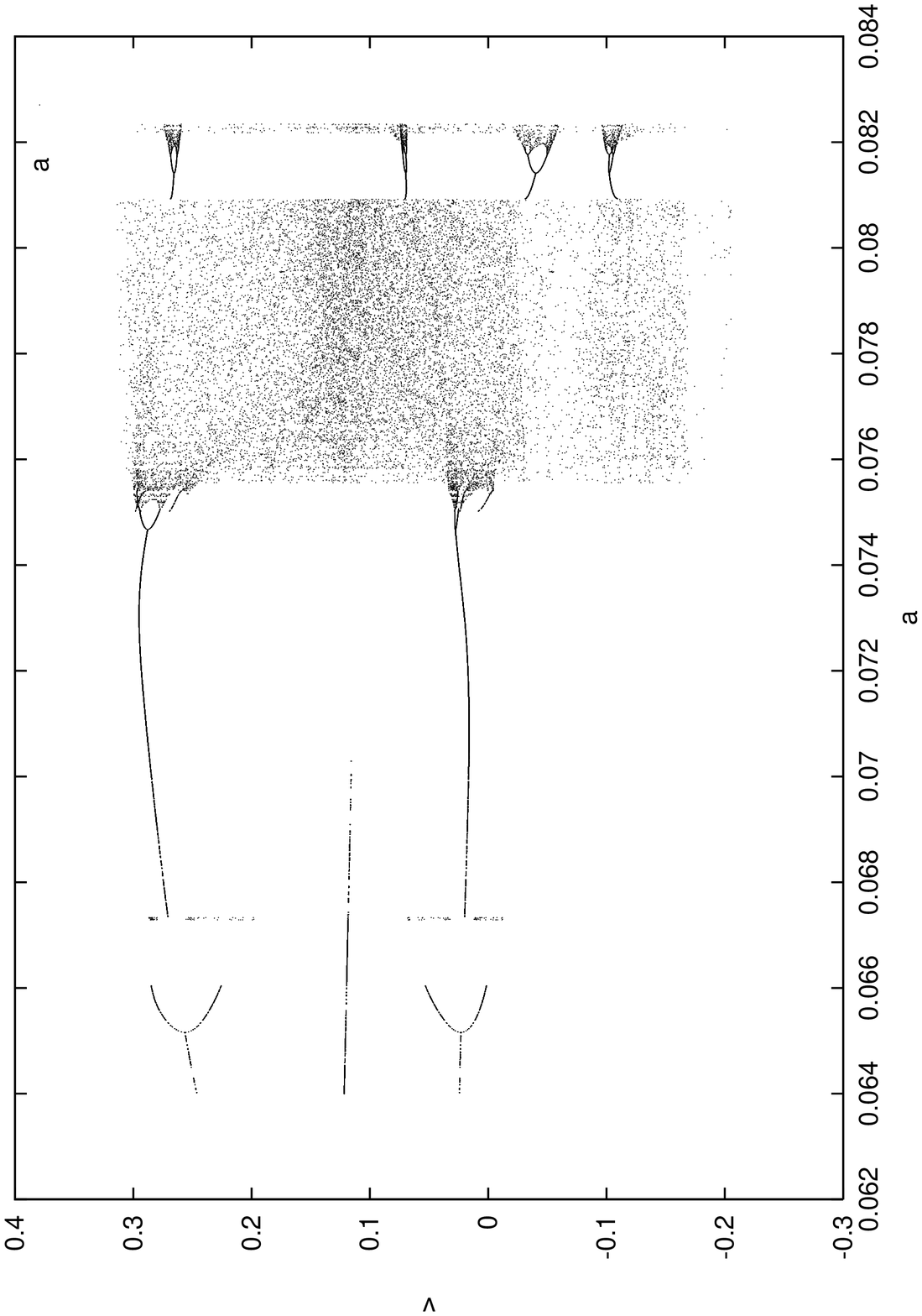,width=8.6cm}}
\label{fig3a}
\end{figure}

\begin{figure}[htb]
\centerline{\epsfig{file=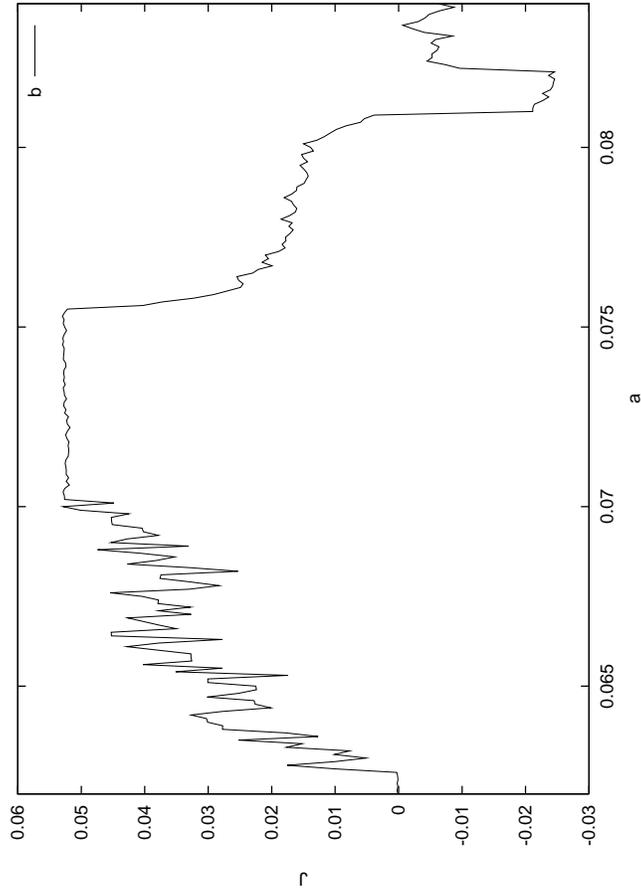,width=8.6cm}}
\caption{For $b=0.1$ and $w=0.67$ we show: (a) The bifurcation
diagram as a function of $a$, and (b) The current $J$ as a
function of $a$. The range in the parameter $a$ corresponds 
to the first current reversal.}
\label{fig3b}
\end{figure}

\vfill\eject

\begin{figure}[htb]
\centerline{\epsfig{file=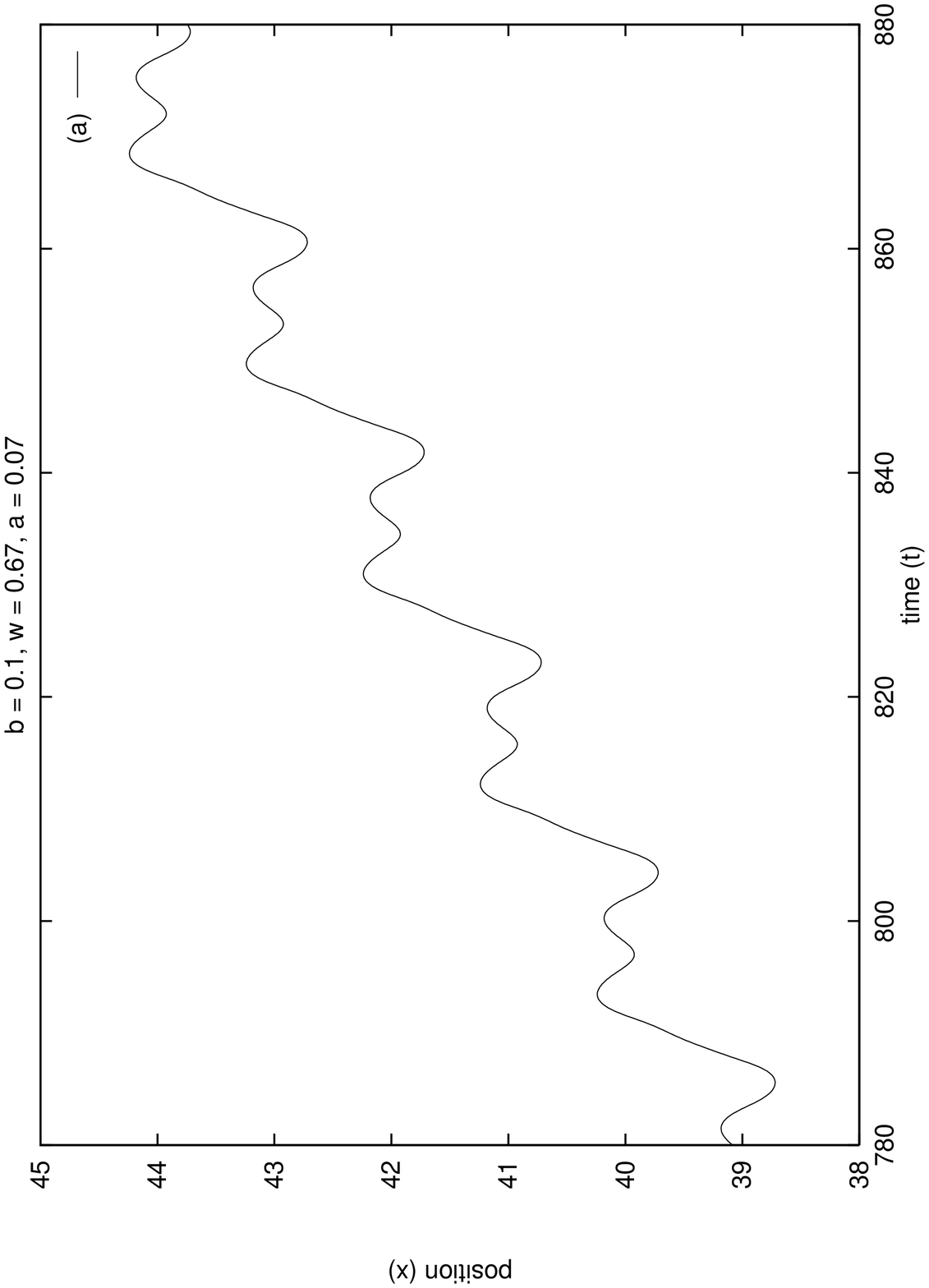,width=7.0cm}}
\label{fig4a}
\end{figure}

\begin{figure}[htb]
\centerline{\epsfig{file=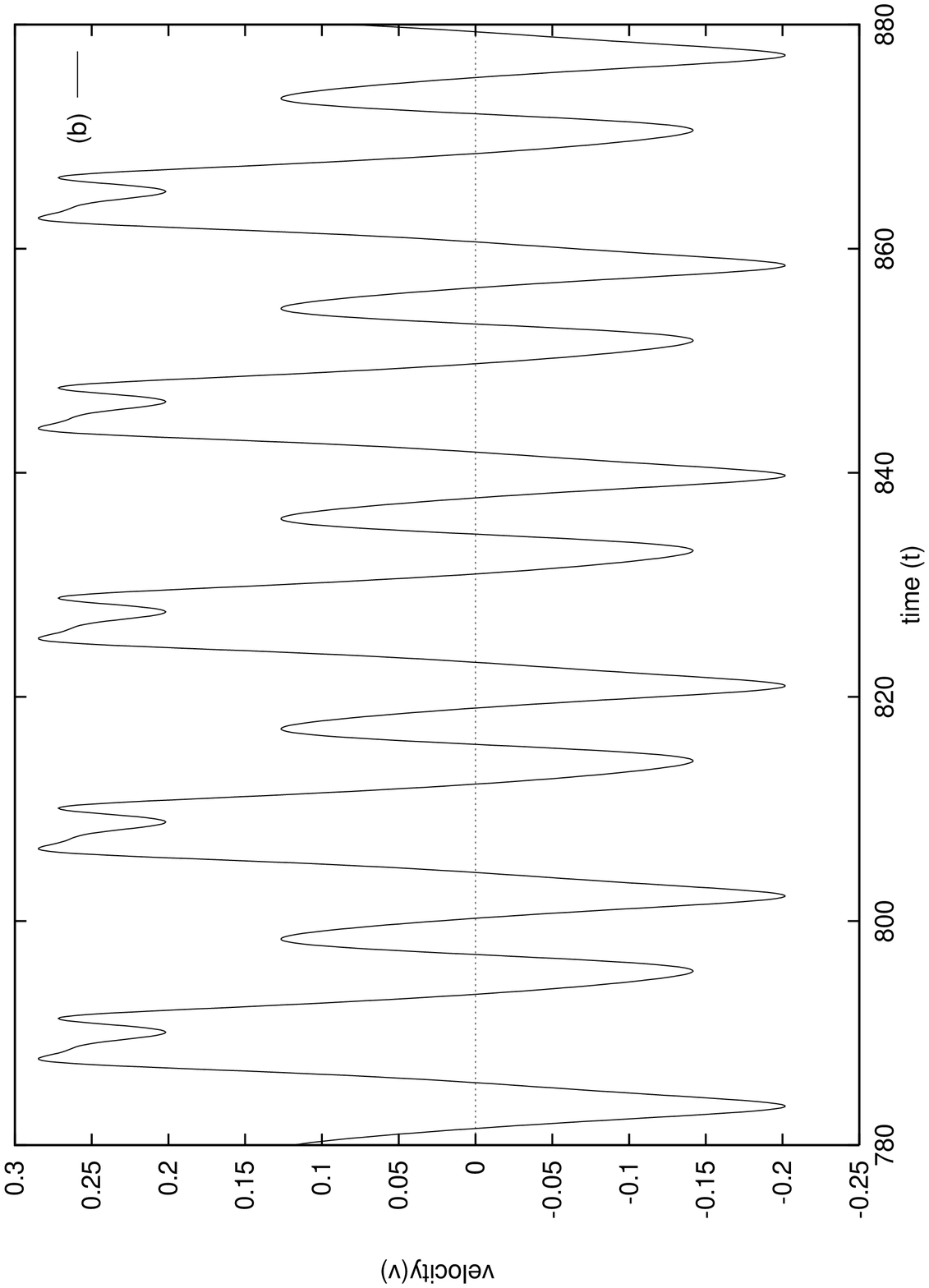,width=7.0cm}}
\label{fig4b}
\end{figure}

\begin{figure}[htb]
\centerline{\epsfig{file=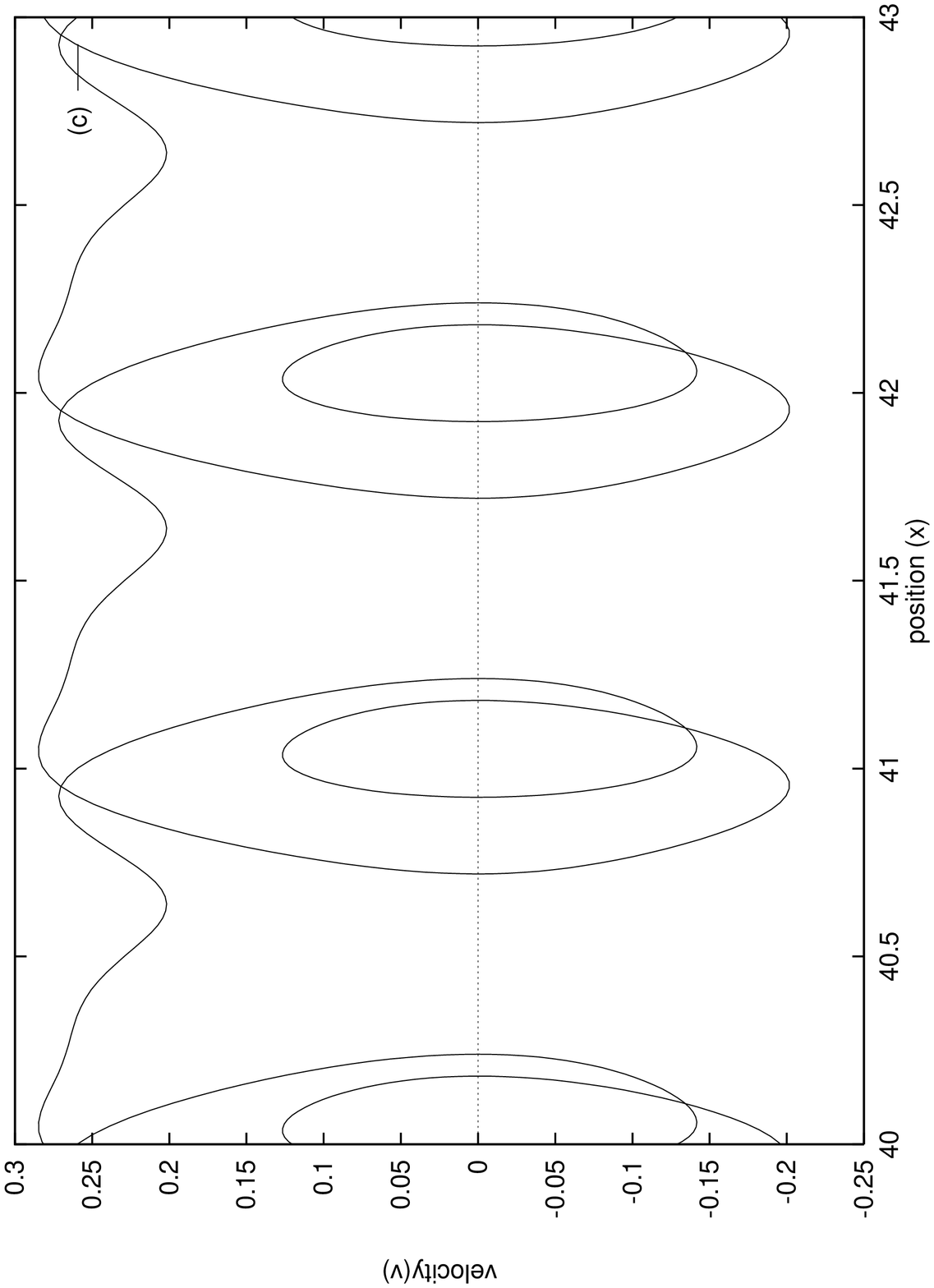,width=7.0cm}}
\caption{For $b=0.1$ and $w=0.67$ and 
$a = 0.07$ we show: (a) The trajectory of the particle 
as a function of time, (b) the velocity as a function of
time and (c) the phase space. This case corresponds 
to positive current.}
\label{fig4c}
\end{figure}

\vfill\eject

\begin{figure}[htb]
\centerline{\epsfig{file=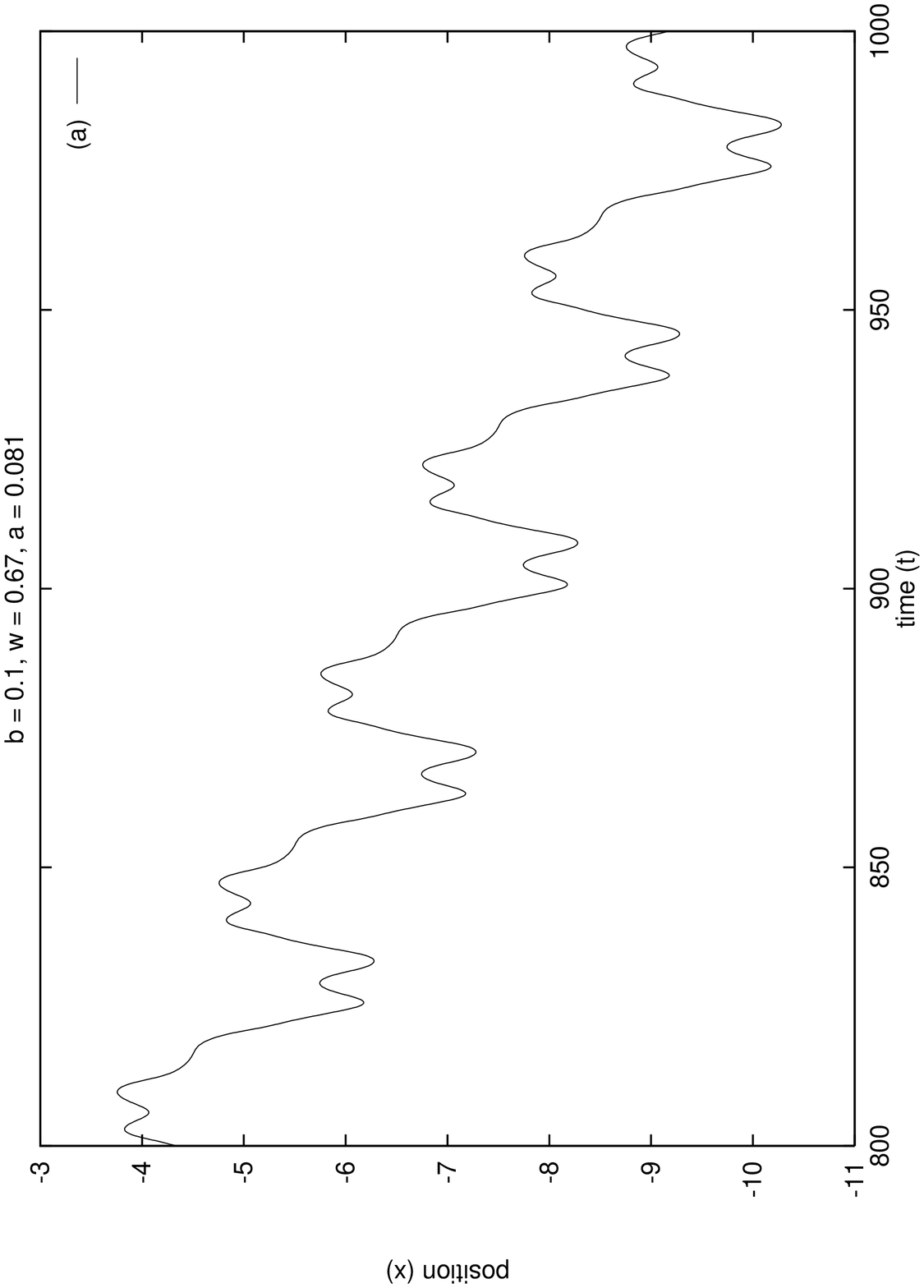,width=7.0cm}}
\label{fig5a}
\end{figure}

\begin{figure}[htb]
\centerline{\epsfig{file=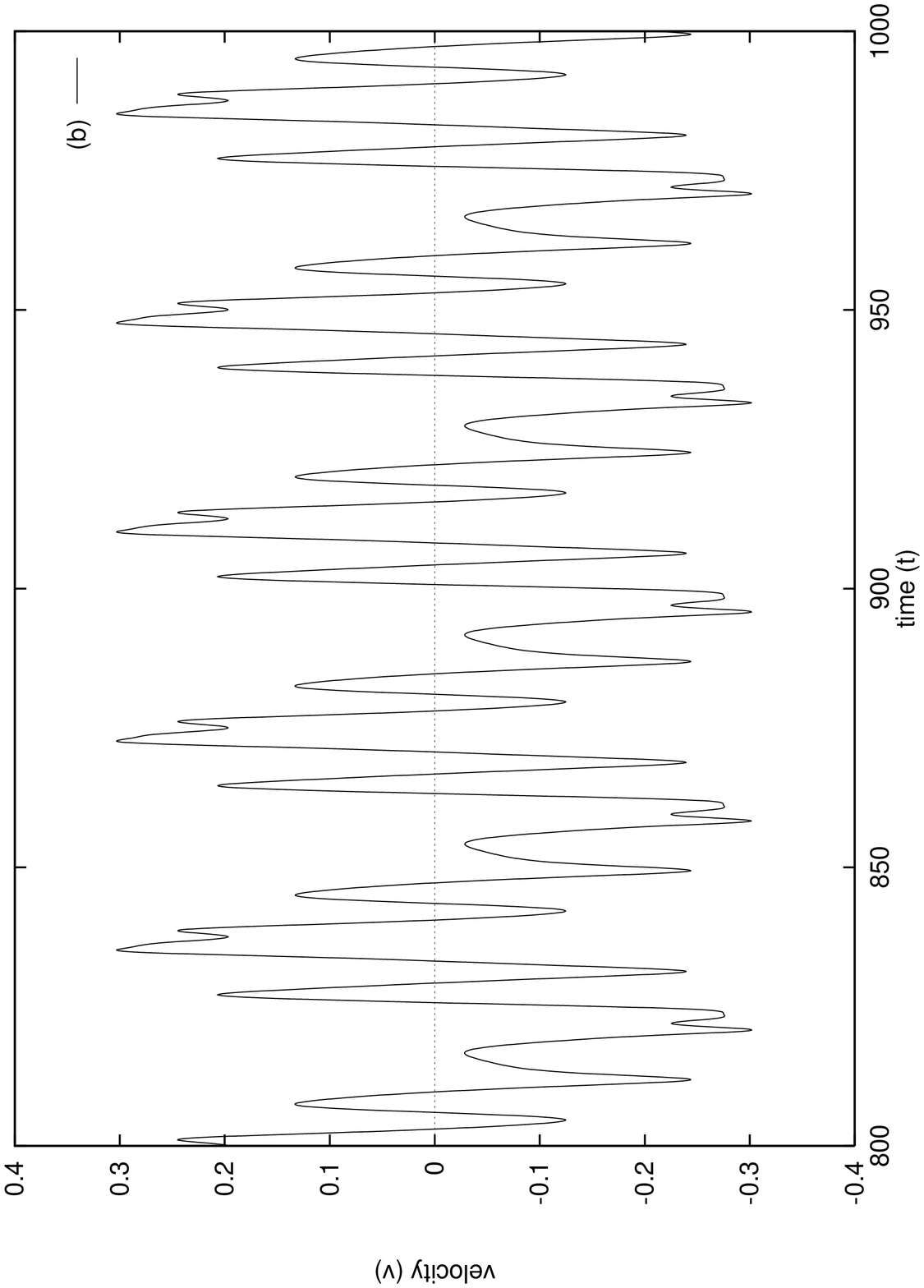,width=7.0cm}}
\label{fig5b}
\end{figure}

\begin{figure}[htb]
\centerline{\epsfig{file=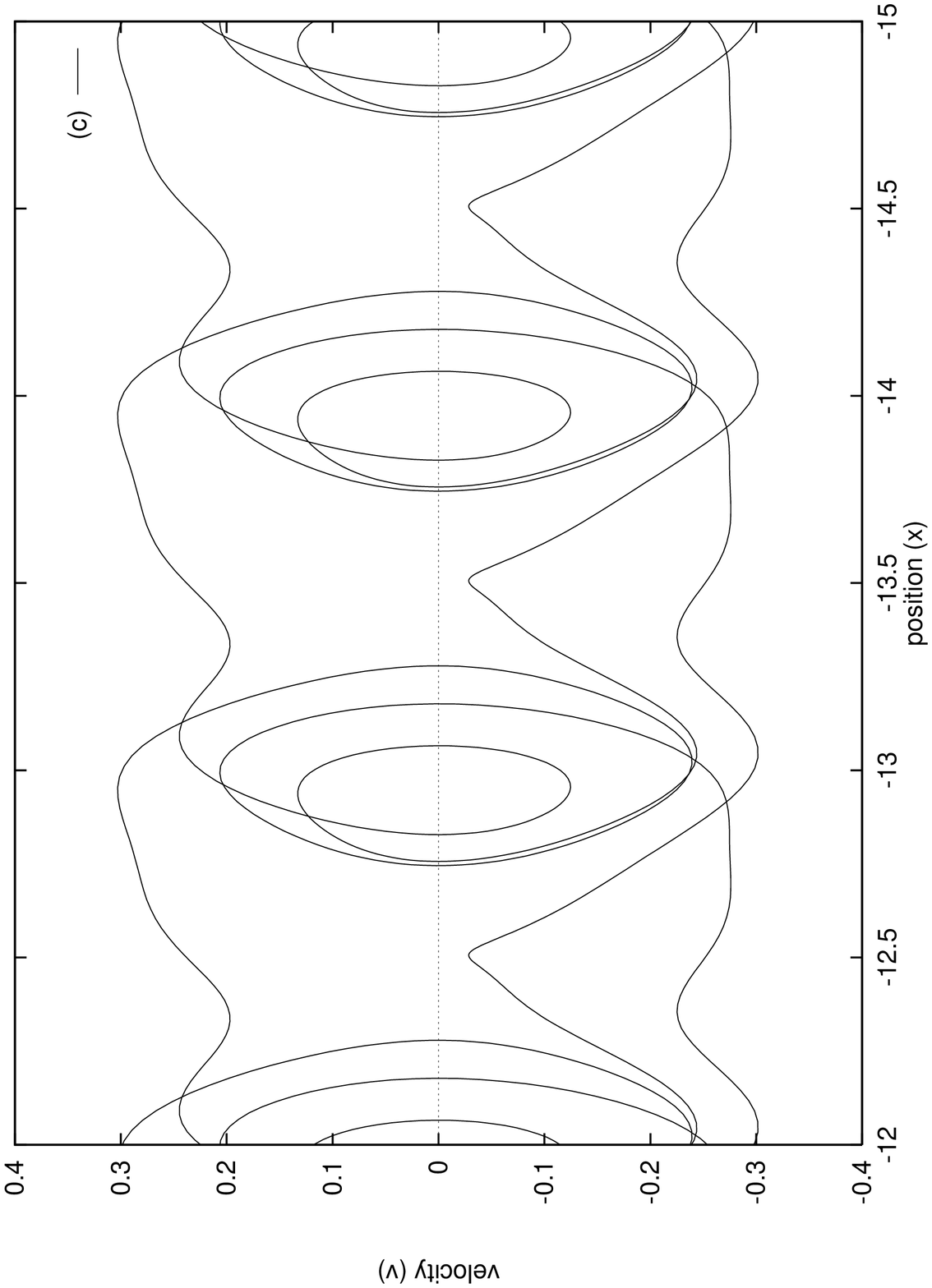,width=7.0cm}}
\caption{For $b=0.1$ and $w=0.67$ and 
$a = 0.081$ we show: (a) The trajectory of the particle 
as a function of time, (b) the velocity as a function of
time and (c) the phase space. This case corresponds 
to negative current.}
\label{fig5c}
\end{figure}

\vfill\eject

\begin{figure}[htb]
\centerline{\epsfig{file=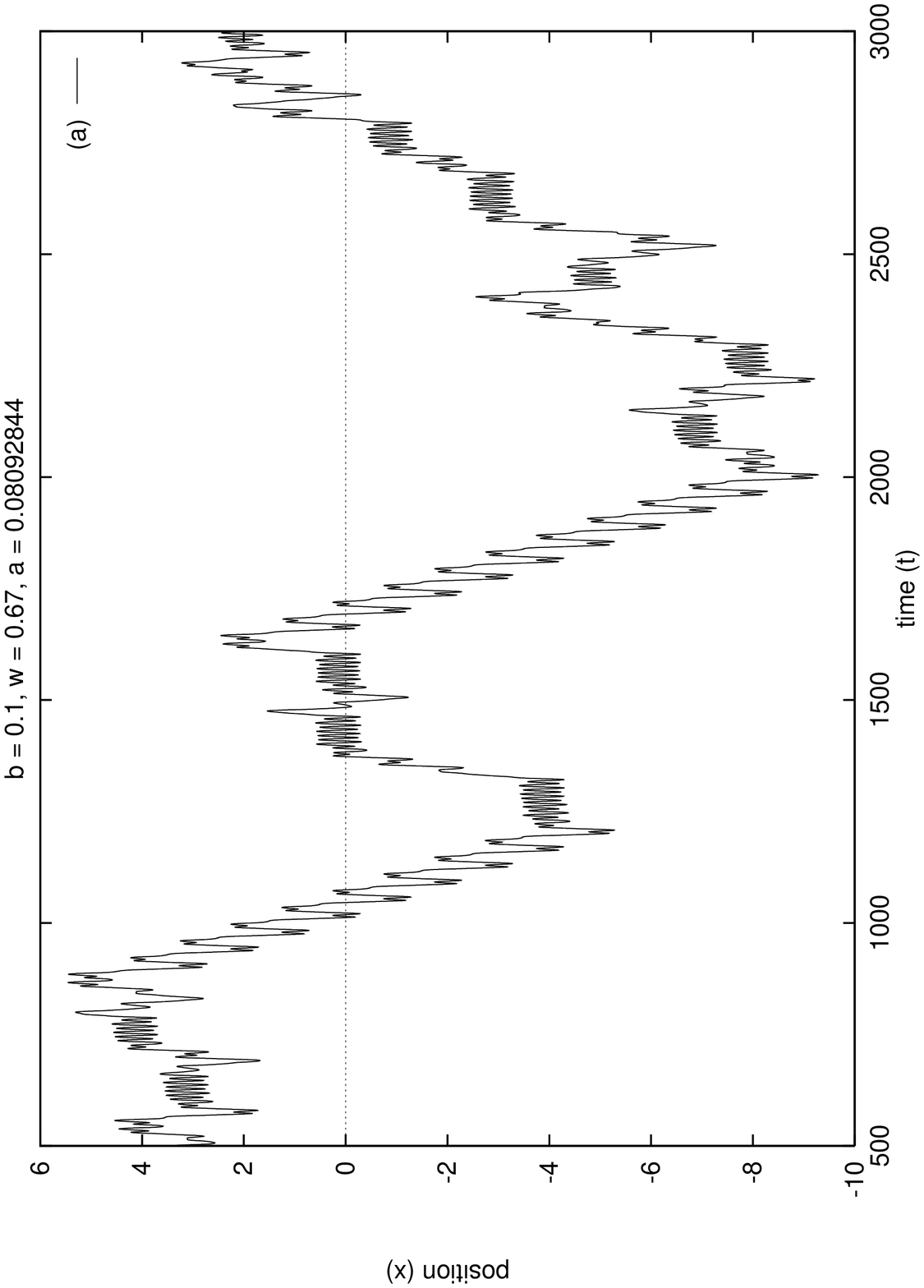,width=7.0cm}}
\label{fig6a}
\end{figure}

\begin{figure}[htb]
\centerline{\epsfig{file=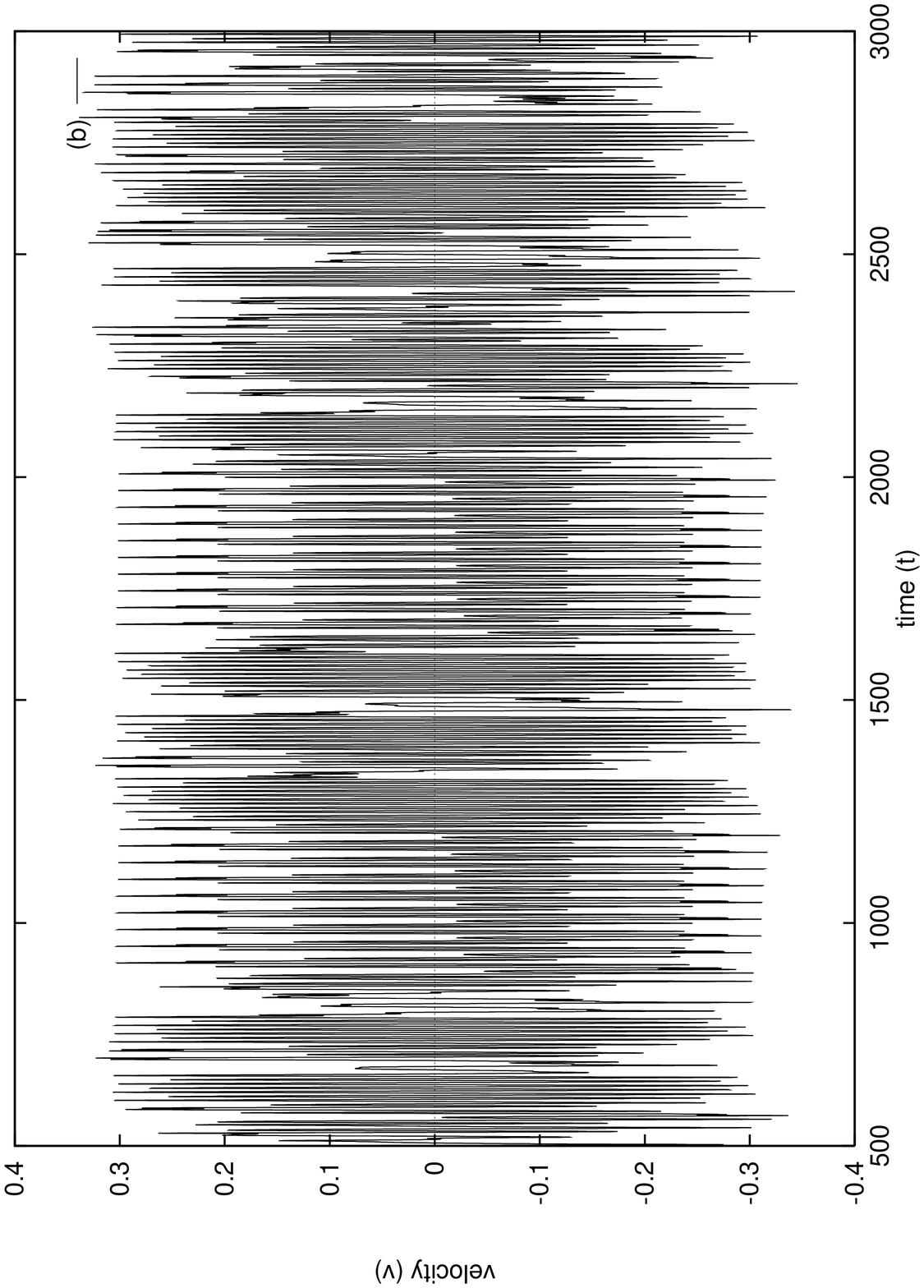,width=7.0cm}}
\label{fig6b}
\end{figure}

\begin{figure}[htb]
\centerline{\epsfig{file=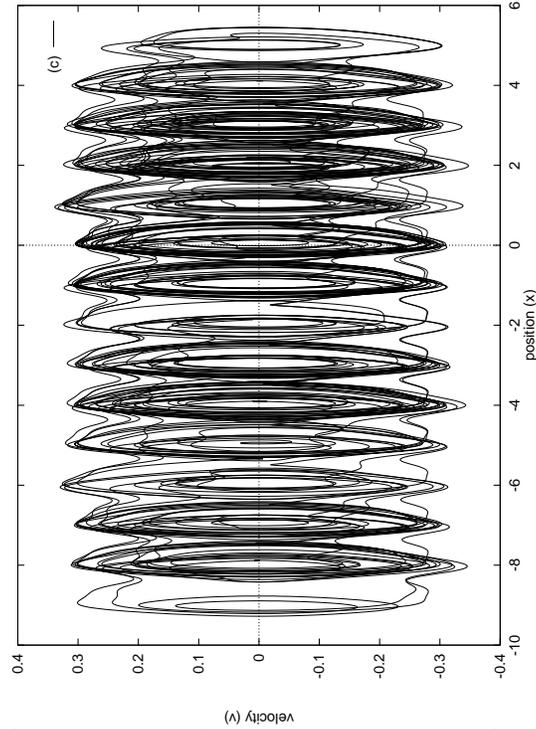,width=7.0cm}}
\caption{For $b=0.1$ and $w=0.67$ and 
$a = 0.080 928 44$ we show: (a) The trajectory of the 
particle as a function of time, (b) the velocity as a 
function of time and (c) the phase space. This case 
corresponds to $a$ near the bifurcation, where the
dynamics becomes intermittent and there is 
anomalous diffusion.}
\label{fig6c}
\end{figure}

\vfill\eject

\begin{figure}[htb]
\centerline{\epsfig{file=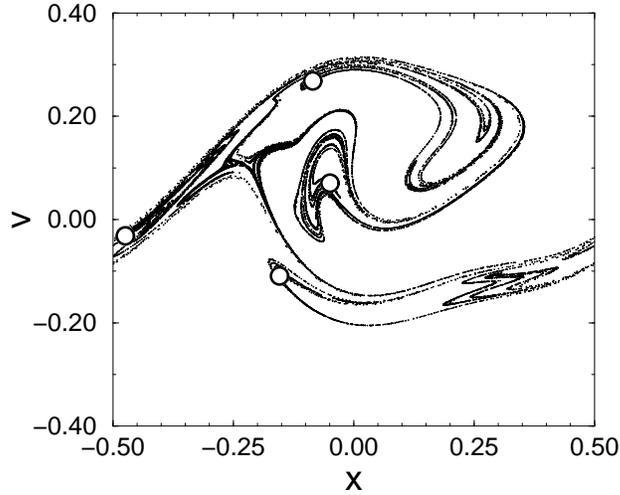,width=8.6cm}}
\caption{For $b=0.1$ and $w=0.67$ we show two attractors: 
a chaotic attractor for $a=0.08092$, just below $a_c$, and 
a period-four attractor, for $a=0.08093$, consisting of 
four points located at the center of the open circles. 
See Ref. [18].}
\label{fig7}
\end{figure}


\end{document}